\documentclass[12pt,preprint]{aastex}

\newcommand{\simlt}{\mbox{$^{<}_{\sim}$}}

\begin{document}
\title{Kinematics of Diffuse Ionized Gas Halos: A Ballistic Model of Halo Rotation}
\author{Joseph A. Collins}
\affil{University of Colorado, Center for Astrophysics \& Space Astronomy, Campus Box 389, Boulder, CO 80309}
\author{Robert A. Benjamin}
\affil{University of Wisconsin, Department of Physics, 1150 University Avenue, Madison, WI 57306}
\and
\author{Richard J. Rand\altaffilmark{1}}
\affil{University of New Mexico, Dept. of Physics and Astronomy, 800 Yale Blvd.
NE, Albuquerque, NM 87131}
\altaffiltext{1}{Visiting Astronomer, National Optical Astronomy Observatories,
Tucson, AZ}

\begin{abstract}

To better understand diffuse ionized gas kinematics and halo 
rotation in spiral galaxies, we have developed a model 
in which clouds are ejected
from the disk and follow ballistic trajectories
through the halo.   The behavior of clouds in this model has been investigated
thoroughly through a parameter space search and a study of individual cloud
orbits.  
Synthetic velocity profiles 
have been generated in $z$ (height above the plane) 
from the models for the purpose of comparing with
velocity centroid data from previously obtained long-slit spectra
of the edge-on spirals NGC 891 (one slit) 
and NGC 5775 (two slits).  In each case, a purely ballistic model
is insufficient in explaining observed DIG kinematics. 
In the case of NGC 891, the observed 
vertical velocity gradient is not as steep as predicted by the model, possibly
suggesting a source of coupling between disk and halo rotation or an outwardly
directed pressure gradient.  The ballistic model 
more successfully explains DIG kinematics observed in NGC 5775; however, it
cannot explain the observed trend of high-z gas velocities nearly reaching
the systemic velocity.  Such behavior can be attributed to either an inwardly
directed pressure gradient or a possible tidal interaction with its 
companion, NGC 5774.
In addition, the ballistic model predicts that clouds move radially outward as 
they cycle through the halo.  The mass and energy fluxes estimated from the 
model suggest this radially 
outward gas migration leads to a redistribution of material that
may significantly affect the evolution of the ISM.

\end{abstract}


\section{Introduction}

The majority of the ionized gas in the interstellar medium of the Milky Way
resides in a vertically extended layer known as the Reynolds Layer or
Warm Ionized Medium (WIM).  Photoionization by massive stars in the disk
is likely the dominant source of energy input into the layer given its
energetic requirements (Reynolds 1993) as well as the first order
agreement between
observed line ratios (Haffner, Reynolds, \& Tufte 
1999) and photoionization models (Domg\"{o}rgen \& Mathis 1994; Sokolowski 
1994; Bland-Hawthorn et al. 1997; Sembach et al. 2000).  

In external galaxies, where the layer is more commonly referred to as diffuse
ionized gas (DIG), early work concentrated on the characterization and 
occurrence of these layers for ``normal'' galaxies (e.g. Rand, Kulkarni,
\& Hester 1990; Dettmar 1992; Ferguson, Wyse, \& Gallagher 1996).
A number of observational
results suggest that local levels of star formation are responsible for
these layers both in energizing the gas and in accelerating the gas upward via
supernova activity and winds. Among the results supporting this conclusion
are the correlation between 
layer prominence (in brightness and extent) and far-infrared luminosity
(e.g. Rand 1996; Rossa \& Dettmar 2000), and with tracers of
star formation at other wavelengths (e.g. Dahlem, Dettmar, \& Hummel 1994).
Further evidence linking DIG halos to star formation
includes the identification of DIG structures 
resembling the chimneys described by Norman \& Ikeuchi (1989), which form 
when expanding supernova-driven supershells break through the main gaseous 
layers of a disk (e.g. Collins et. al. 2000). 
In recent years, the edge-ons NGC 891 and NGC 5775 have provided a wealth of
information on the origin and excitation of gaseous halos.  Each of these 
galaxies has relatively high rates of star formation as indicated by their
far-infrared luminosities.  NGC 5775 is classified as a starburst galaxy
(Condon \& Broderick 1988), though there is little evidence for a bright, 
nuclear starburst.  DIG halos in these
galaxies are considerably more extended in $z$ than the Reynolds Layer, and
in the case of NGC 5775, DIG can be detected up to $z\approx13$ kpc (Rand 
2000).  In each
of these galaxies, line ratio data indicate that a non-negligible fraction of 
DIG halo emission originates in a component that is either
energized by an additional ionizing source such as shocks (e.g. Shull \& McKee
1979) or
is at a considerably higher temperature (e.g. Reynolds, Haffner, \& Tufte
1999) than the bulk of halo DIG
(Rand 1998; T\"{u}llman et.
al. 2000; Collins \& Rand 2001). 

Though significant research has been carried out regarding their
energization, relatively little work has been pursued on the issue of
rotation and support of these vertically extended layers through
studies of DIG kinematics.  Current evidence suggests that halo gas
rotates more slowly than gas in the disk.  In the case of NGC 891, the
HI halo component appears to rotate 25-100 km s$^{-1}$ more slowly
than the disk component (Swaters et. al. 1997).  In addition, the ``bearded''
HI position-velocity map of the intermediately inclined
galaxy NGC 2403 has been successfully modeled as a superposition of a
thin disk and a more slowly rotating thick halo (Schaap et. al. 2000;
Fraternali et al. 2001).  
At higher-redshift, halo kinematics has been 
investigated  through quasar absorption-line studies (e.g. Charlton \& 
Churchill 1998).  The recent results of Steidel et al. (2002) are 
consistent with a kinematic model where the QSO-absorbing gas resides in a 
halo which rotates more slowly than its parent galaxy.
     
However, though HI data cubes yield information for gas close to the
midplane, they shed little light on gas 
kinematics at highest $z$ for low-redshift galaxies,
which is better traced by the more extended ionized gas layer.  In NGC
891 (Rand 1997) and NGC 5775 (Rand 2000; T\"{u}llmann et. al. 2000),
emission line centroids indicate a slow-down in DIG rotation with $z$.
However, in the case of NGC 891, Benjamin (2000) has found that the
observed drop-off in rotation speed from $z=1$ to 5 kpc of $\sim30$ km
s$^{-1}$ is considerably more shallow than the value of 80 km s$^{-1}$
determined by assuming a fluid disk in hydrostatic equilibrium and
calculating rotation speeds in a standard galactic potential.  This
calculated value is only weakly dependent on the choice of potential
and is predominantly due to the geometric effect of a decrease in the
projection of the radial gravitation vector with height above the
disk.  Higher order effects included in hydrodynamical models of halo
kinematics, such as outwardly directed pressure gradients or a coupling
between disk and halo through magnetic tension or viscosity, may
resolve this discrepancy.  However, in practice these effects are not
well constrained given the lack of understanding on issues such as
galactic magnetic field structure and gas densities.

Though a fluid disk in hydrostatic equilibrium is perhaps the simplest
model one can adopt in considering halo kinematics, it assumes that
the observed gas is dynamically coupled to the surrounding gas. This
is not necessarily the case.  An alternate point of view is that
extraplanar gas consists of density concentrations sufficiently large
that the motion is essentially ballistic.  These clouds then cycle
through the halo similar to Bregman's (1980) galactic fountain return
flow. Among the pieces of evidence that support such a picture is the
fact that DIG layers are considerably thicker (by more than an order
of magnitude) than the thermal scale height of 10$^{4}$ K gas in the
gravitational field (e.g. Dettmar 1992) indicating turbulent motion
and outflows, or, at the very least, a source of pressure support to
maintain these thick layers.  This fact, coupled with strong
evidence of outflows in normal galaxies (e.g. Golla \& Hummel
1994; Collins et al. 2000), indicate extensive cycling of material
from disk to halo.  Recent observations of a quasar projected at a
height of $z=5$ kpc behind the halo of NGC 891 indicate a filling
factor of low-ionization species (\ion{Mg}{2}\ and \ion{Fe}{2}) gas of
around $f\sim0.025$ (E. Miller, private communication).  This suggests
a very clumpy medium if in fact this gas traces the DIG component, and
lends some justification to treating the ISM as a collection of
circulating clouds.

Towards the goal of a better understanding of DIG kinematics and halo 
rotation, we have developed an ISM model which assumes DIG is distributed
in clouds which take ballistic trajectories as they move from disk to halo 
and back.  We then attempt to compare velocity 
centroid data, obtained from previously presented long-slit data for NGC 891 
(Rand 1997) and NGC 5775 (Rand 2000), with this ballistic cloud model.  In all
cases the slit is oriented perpendicular to the midplane so that these issues 
can be addressed as a function of $z$. Further details of the observations
can be found in the relevant references.  The ballistic model is discussed 
in \S\ 2.  Data are compared to the model and results are discussed in \S\ 3.
Conclusions are summarized in \S\ 4.

\section{Ballistic Models of Gas Kinematics}

In order to study how halo gas should rotate we have constructed a 
model to calculate cloud orbits for a given set of initial conditions.  Using 
the potential model of Wolfire et. al. (1995), we calculate $x, y$, and $z$ 
velocities for 2$\times10^{4}$ clouds as they move from 
disk to halo and back again
in this potential.  The velocity at the flat part of the rotation curve, 
$V_{c}$, is a free parameter that can be chosen to vary the potential 
according to the particular galaxy we wish to model.  For NGC 891 and NGC 5775
we set $V_{c}$ to HI rotational velocities from the literature 
for each of these 
galaxies: $V_{rot}=230$ km s$^{-1}$ (Swaters et. al. 1997) and $V_{rot}=198$
km s$^{-1}$ (Irwin 1994), respectively. 

We assume each cloud is initially located in the  
disk at a location determined by probability distributions 
in $R$ and $z$:

\begin{equation}
P(R)\propto e^{-R/R_{0}}
\end{equation}
and
\begin{equation}
P(z)\propto e^{-z^{2}/2s^{2}}, s=0.2 \ \mbox{kpc}.
\end{equation}

These H$\alpha$ disks have very sharp radial cut-offs. Thus to 
best match observational data, the radial scale length of the initial 
cloud distribution, $R_{0}$, is chosen to be one-half the radius of the 
cut-off in the 
H$\alpha$ emitting disk.  From the H$\alpha$ images we have set $R_{0}=6$ kpc 
for NGC 5775 (Collins et. al. 2000) and $R_{0}=7$ kpc for NGC 891 (Rand, 
Kulkarni, \& Hester 1990).  Each cloud is given a random vertical 
``kick'' velocity in the $z$ direction, the probability distribution for 
which is uniform between zero and 
a chosen maximum kick velocity, $V_{k}$, to 
simulate the initiation of a disk-halo cycle.  In addition, the clouds are 
ejected at an angle, $\gamma$, from the vertical determined by the probability
distribution:

\begin{equation}
P(\gamma)\propto e^{-\gamma^2/2\gamma_{0}^{2}}.
\end{equation}

We assume these clouds are non-interacting (no drag) and remain fully ionized
throughout their orbits.  We thus neglect any phase change that may occur such 
as conversion to HI in the halo. 
This model is therefore similar to the ``cannonball'' models 
considered by Wakker (1990) and Charlton \& Salpeter (1989).  
This assumption is a 
major source of uncertainty in these models as hot and warm gas ejected from
the disk is thought to be a source of the infalling high-velocity clouds 
(HVCs) observed in \ion{H}{1}\ near the Milky Way 
(e.g. Wakker \& van Woerden 1997).   However, we have generated profiles 
assuming the clouds are ionized only during the upward part of their trajectory
and find that any changes in mean velocity vs. $z$ are insignificant.
When any of the 2$\times10^{4}$ clouds returns to the midplane, a new cloud is 
put into play. After 
1 Gyr, the sample of clouds is relaxed to an approximate steady state and 
spatial and velocity information for each cloud is extracted.  Since we 
are considering a case without drag, these resulting values are independent of
initial cloud mass, density, and temperature  which are thus taken to be 
equal for each cloud.  In addition, we make the further assumption that the 
density and temperature of each cloud does not change during orbit.  With 
these assumptions the total H$\alpha$ flux for any line of sight through the 
model is then simply proportional to the number of clouds intercepted by that 
line of sight.  

The model is then placed at an inclination indicated by the HI data 
[$i\approx90\arcdeg$ for NGC 891 (Swaters et. al. 1997) and $i=86\arcdeg$ 
for NGC 5775 (Irwin 1994)]
and cloud properties are calculated for bins of 1 kpc extent along the major 
axis. Line of sight velocities are then
averaged in 1 kpc bins in $z$ to generate a synthetic velocity profile versus
$z$.  These velocities represent mainly rotation, though line of sight 
contributions from radial and, in the case of the not quite fully edge-on
NGC 5775, vertical motions are included as well. 

\subsection{Parameter Space Search of Ballistic Models}

To better understand the behavior of the model we have performed a restricted
search of parameter space around what will be reported 
to be the best models for
NGC 891 and NGC 5775. The free parameters that determine the behavior of cloud 
orbits are the maximum kick velocity, $V_{k}$, the circular velocity, $V_{c}$,
the ejection cone angle, $\gamma_{0}$, the initial scale height of clouds,
$s$, and the initial radial distribution of clouds.  
The radial scale length of the 
initial cloud distribution, $R_{0}$, is constrained by observations. However,
we can still adjust the radial distribution by including a central hole to
the cloud distribution described by the parameter $R_{hole}$.  The circular
velocity, $V_{c}$, is also directly constrained by \ion{H}{1}\ rotation curves.
In addition, we find that the initial scale height of the cloud distribution, 
$s$, has an insignificant effect on model outputs for values appropriate for 
an initially thin gas disk.  Thus the parameters affecting model outputs 
which are not directly constrained by observations are $V_{k}$, 
$\gamma_{0}$, and $R_{hole}$. The number of clouds, observed heliocentric 
velocity, and velocity dispersion versus $z$ in models of NGC 891 and 
NGC 5775 are shown in Figures 1 and 2  at
slit positions of $R=0$, 4, and 8 kpc along the major axis for a variety
of different combinations of parameters. The main 
situations we wish to explore are the effect of the kick velocity, $V_{k}$, 
the inclusion of a cone angle, $\gamma_{0}$, and the effect of a central
hole in the initial cloud distribution. 

One can immediately see some general trends common to the various models shown
in Figures 1 and 2.  First, each of the models 
produces a cloud distribution that is approximately exponential in $z$.  
Second, the various models all predict a decrease in mean
heliocentric velocity with $z$ due to decreased rotation speeds in the 
halo.  In the case of NGC 5775, the run of $V_{hel}$ vs. $z$ is asymmetric
about $z=0$ due to the fact that the galaxy is not fully 
edge-on.  As a result $V_{hel}$ includes contributions from
not only the rotation velocity, but vertical and radial velocities as well.
Finally, the line widths of the models each exhibit peaks about $z=0$ kpc.  
This peak   
for $R=0$ kpc is particularly sharp and is due mainly to the fact that we are 
using 1 kpc bins along the major axis.  Thus this bin includes clouds 
moving parallel and perpendicular to the line of sight near the galactic 
center.  At other values of $R$, the high velocity dispersion at low-$z$ 
is due to contributions from both clouds that have recently been 
ejected from the disk (and thus still having relatively high rotation 
speeds) and clouds that are falling back to the disk (and thus rotating more
slowly).  

The maximum kick velocity,
$V_{k}$, is a key parameter in determining how high into the halo these 
clouds can reach.  For each galaxy, we initially choose $V_{k}$ such that 
the scale
height of the cloud distribution, $h_{cl}$, matches the scale height of the
bulk of the emission from the spectrum of NGC 891 of Rand (1997)
and the scale height determined for NGC 5775 from Collins et al. (2001).
For NGC 891 we set $V_{k}=100$ km s$^{-1}$ to generate
$h_{cl}=1$ kpc, while for NGC 5775  values of $h_{cl}=2.2$ kpc and 
$h_{cl}=2.1$ kpc at the radial positions of Slits 1 and 2 are obtained by
setting $V_{k}=160$ km s$^{-1}$.  These particular models, called the ``base 
models,'' are fully specified in Tables 1 and 2.
The base models for NGC 891 and NGC 5775 are shown in Figures 
1 and 2, along with models with slightly higher 
and lower kick velocities.   
Scale height, 
the change in $V_{hel}$ from the midplane to $z=3$ kpc, and the range of 
$\sigma_{v}$ in the various models are shown in Tables 1 and 2. 
Clearly the most significant effect 
of increasing the kick velocity is to increase the scale height of the cloud
distribution as well as the line widths.  The effect on the mean velocity 
profile, however, is more subtle with smaller kick velocities producing a 
slightly steeper fall-off with $z$.

\placetable{t1}

\placetable{t2}

In Figures 1 and 2 we have also plotted a model where a central hole in the
initial cloud distribution has been introduced to the base model.  The radius
of the hole is chosen to be one-half the radial scale length of the initial
cloud distribution ($R_{hole}=3.5$ kpc and $R_{hole}=3$ kpc for NGC 891 and 
NGC 5775, respectively).  The introduction of a central hole has
very little effect on the model outputs.  The exception is when the slit 
position line-of-sight crosses the central hole such as for $R=0$ kpc.  In 
this case, the mean velocity and dispersion do not include clouds moving
parallel to the line of sight and are therefore lower. 

In Figures 1 and 2, a model is plotted where a cone angle of 
$\gamma_{0}=20\arcdeg$ is introduced to the base model.  The inclusion
of the cone angle has a negligible effect on scale height and mean velocities
versus $z$.  The main effect of introducing a cone angle is to increase the 
line widths over that of the base model.  Clearly, determinations 
of line widths
for these slits would help distinguish the appropriate choice of cone angle
for these models. However, the spectral resolution of our data is inadequate
for such measurements. 

In summary, the parameter space search reveals that these models show 
surprisingly little variation in the predicted mean velocity profile for any 
reasonable set of input parameters.  Clearly the most important parameters 
affecting the profiles are the circular velocity, $V_{c}$, which determines 
the strength of the potential, and the kick velocity, $V_{k}$.  Even the 
effect of significantly varying $V_{k}$ on the velocity profiles is somewhat 
small considering its effect on the scale height.  As a result, we can be 
fairly certain that any uncertainty in model input parameters does not
significantly affect the conclusions drawn from our comparisons in \S\ 3.

\subsection{Behavior of Individual Orbits}

In order to gain some insight into the trends presented in the previous
section, we also calculated a grid of individual orbits in order to
understand what factors determine the heights, velocities, and radial
motion of the orbits. In the cases we initially consider, we have used
the gravitational potential used in Wolfire et al (1995). For this
form of the potential, each component (bulge, disk, and halo) of the
gravitational potential function scales linearly with $V_{c}^2$. As a
result, we can write the equation of motion for a cloud as

\begin{equation}
\frac{d\tilde{V}}{d(V_{c}t)}={\bf \tilde{g}}(R,z),
\end{equation}
where $\tilde{V}=V/V_{c}$ is the normalized cloud velocity, and ${\bf
\tilde{g}}={\bf g}/V_{c}^{2}$ is the normalized gravitational
acceleration. If an orbit is initialized with some kick velocity
$V_{k}$, those orbits with the same value of $V_{k}/V_{c}$ will
have the same physical path. The time it takes to complete the orbit
will
scale inversely with the circular speed characterizing the
galactic potential.

In Figure 3, we show four sample orbits up to the first plane
crossing.  These orbits are characterized by a nearly vertical climb,
followed by a radially outward drift and descent back to the plane. In
all cases, the orbits intersect the galactic plane at a larger radius
than their initial coordinates, meaning that vertical galactic
circulation is primarily a source of outward mass and momentum
transport. The rotation velocity at any radius can be derived using
conservation of angular momentum, so that $rV_{\phi}=$ constant. The
average outward radial velocity shown in Figure 4, calculated by dividing
the radial distance traveled by the total time to cross the midplane, 
is relatively insensitive
to $V_{k}/V_{c}$ and initial radius, lying in the range 15-30 ${\rm
km~s^{-1}}$ when $V_{k}/V_{c}> 0.5$ and $R_{0}>5$ kpc ($R_{0}$ denotes
the orbit's initial galactocentric radius), assuming $V_{c}=200$ km s$^{-1}$. 
For smaller $R_{0}$, the average 
radial velocity exceeds 30 km s$^{-1}$ when $V_{k}/V_{c}> 1$.   
For $V_{k}/V_{c}< 0.5$ the value drops below 10 km s$^{-1}$, although  
there is some dependence on $R_{0}$. 
This average outward radial velocity scales 
linearly with the circular velocity. For orbits originating at large $R_{0}$,
there is some
slight radially inward motion just before the orbit crosses the
plane. Such radially inward motion has been claimed to have been
detected by Fraternali et al. (2001) in the galaxy NGC 2403.  The
maximum height of the orbit and time to cross the midplane depend on
both the initial radius, $R_{0}$ and $V_{k}/V_{c}$ are also shown in
Figure 4.

Figure 4 also presents the outward radial migration as a function of
initial radius and $V_{k}/V_{c}$.  For initial radii greater than
$R_{o}=4$ kpc (i.e. away from the bulge), the fractional change in
radius seems to be a function of $V_{k}/V_{c}$ only, increasing from
10\% at $V_{k}/V_{c} \cong 0.4$ to 70\% at $V_{k}/V_{c}=1.0$ with a
roughly linear dependence over this range. This behavior makes it
easier to estimate the outward mass flux analytically. We have also
calculated a grid of orbits using different forms for the galactic
potential: models 2 and 2i of Dehnen \& Binney (1998). The resulting
radial drift curves are shown in Figure 5.  For $V_{k}/V_{c} \simlt
1$, Model 2 produces a similar result as the Wolfire model, but results
in greater radial drift than for the Wolfire et al. potential for
large values of $V_{k}/V_{c}$. Model 2i, however, has an oblate
flattened halo with axial ratio $q=0.3$, (as opposed to $q=0.8$, see
Dehnen \& Binney (1998) for more details) and exhibits a much steeper
rise in radial drift with $V_{k}/V_{c}$. These curves should be useful
in modeling the long term effects of surface mass density and angular
momentum of the galactic disk due to vertical circulation; such models
must also specify how the momentum is transferred back to the disk via
drag forces and possibly include the effects of radial inflow in the
midplane.

\subsection{Mass and Energy Fluxes}

Since the distribution of clouds that we calculate reaches a
steady-state, a given halo configuration requires a certain mass flux
to sustain it. The orbits presented here are independent of the mass
of the clouds. However, if one prescribes the desired total mass of
the halo circulations, our calculations prescribe the mass flux. This
is given simply by $\dot{M_{h}}= M_{h} (\dot{N}/N)$ where $M_{h}$ 
is the halo mass
and $N$ is the number of clouds in our simulation. The cycling frequency
$f_{cycle}=\dot{N}/N$, the inverse of the mean timescale for the
circulation, is given by our calculation and depends on the initial
probability distribution functions presented at the beginning of
Section 2. Keeping these values fixed but varying $V_{k}/V_{c}$, we
find that 

\begin{equation}
\dot{M_{h}} ( V_{k}/V_{c}) =f_{cycle} M_{h}~ ({\rm M_{\sun}~
yr^{-1}}), 
\end{equation}
where $M_{h}$ is the total halo mass in solar masses.
Values of the cycling
frequency, $f_{cycle}$, for two different forms of the galactic
potential, are given in Figure 6. NGC 891, with $V_{k}/V_{c}=0.435$, has
$f_{cycle}=2.0 \times 10^{-8}~ {\rm yr^{-1}}$, while NGC 5775, with
$V_{k}/V_{c}=0.808$, has $f_{cycle}=1.4 \times 10^{-8}~ {\rm yr^{-1}}$.

To determine the mass circulation rates inferred from these models, we
must therefore estimate the halo mass.  We use the H$\alpha$ images of
NGC 891 (Rand et al. 1990) and NGC 5775 (Collins et al. 2000) to
determine the halo masses.  We assume the emitting gas is concentrated in
clouds of constant density as in the model. 
With this assumptions, the emission measure from the H$\alpha$ image can
then be converted to electron column density along the line of sight:

\begin{equation}
N_{e}(\mbox{cm$^{-2}$})=\left(4.37\times10^{19} \sqrt{\frac{f_{V}}{0.2}}\right) \sqrt{EM{\cdot}L},
\end{equation}
where the emission measure, $EM=\int n_{e}^{2} dl$, is in units of pc 
cm$^{-6}$
(we use a conversion from H$\alpha$ intensity of
$2.0\times10^{-18}$ erg s$^{-1}$ cm$^{-2}$ arcsec$^{-2}=1$ pc cm$^{-6}$  
assuming a 10$^{4}$ K gas), $L$ is the path length through the halo
in units of kpc, and $f_{V}$ is the volume filling fraction of the clouds.
We leave the expression in terms of $f_{v}/0.2$, where 0.2 is the approximate
filling factor of the Reynolds Layer (Reynolds 1993).  
We consider halo emission only, as the disk is dominated by bright 
\ion{H}{2}\ regions.  The H$\alpha$ emitting halo mass can then be calculated:

\begin{equation}
M_{h}({\rm M_{\sun}})=\left(8.01\times10^{-15} \sqrt{\frac{f_{V}}{0.2}}\right) \sum_{halo}N_{e}{\Delta}s^{2},
\end{equation}
where ${\Delta}s^{2}$ is the pixel size in kpc$^{2}$.

The estimated halo masses and halo mass fluxes for NGC 891 and NGC 5775 
are shown in Table 3.  The halo mass of NGC 891 is nearly a factor of
three greater than the value estimated by Dettmar (1990), though the scale
height of H$\alpha$ emission used for that calculation ($h_{em}=500$ pc) is 
somewhat lower than indicated by our data.  We note that the estimated mass
fluxes are highly uncertain due to the uncertainties in the calculation of halo
masses.    

\placetable{t3}

These mass fluxes are somewhat higher than expected for these galaxies, for
a filling factor of 0.2,
indicating that a ballistic model is perhaps a too simplistic model.  
The ballistic model, in effect, assumes all extraplanar gas is in the process 
of cycling through the halo in an orbit as described in \S\ 2.2.  In contrast,
gas in a purely hydrostatic fluid disk model does not circulate.  
The ballistic model
thus maximizes the mass flux for a given halo mass and, as a result, the values
calculated above should be thought of as upper limits to the actual flux.

We also note that we may use this model to estimate the kinetic energy flux
associated with the circulation, given by $\dot{E}= (3.2 \times
10^{39}~ {\rm ergs~ s^{-1}}) \dot{M_{h}} <v_{k,100}^{2}>$ where the mass
flux is in ${\rm M_{\sun}~ yr^{-1}}$ and the kick velocity is in units
of ${\rm 100~km~s^{-1}}$.  For our assumed kick velocity distribution
function, which is constant over the range $v_{k}=0$ to $v_{k}=V_{k}$,
$<v_{k,100}^2>=1/3V_{k,100}^{2}$, so that $\dot{E}= (1.1 \times
10^{39}~ {\rm ergs~s^{-1}}) \dot{M_{h}} V_{k,100}^{2}$.  The calculated halo
kinetic energy fluxes for NGC 891 and NGC 5775 are shown in Table 3.
Note that this energy flux is well below the canonical
$\dot{E}=10^{42}~{\rm ergs~ s^{-1}}$ assuming one supernova
every 30 years for a Milky Way-like spiral galaxy. Thus this
circulation does not violate any energetics requirement.

\section{Comparisons Between Data and Models}

As stated earlier, we call the model that best matches the observed scale 
height of emission for each galaxy the base model.
The parameters used in the base models of NGC 891 and NGC 5775
are listed  in Tables 1 and 2, respectively.  Models with non-zero cone
angle and  central holes in the initial gas 
distribution, though able to match the observed scale heights, cannot be well
constrained from available data in that these parameters have negligible 
effects on the generated profiles.  These parameters could be constrained with
comparisons to line-widths; however, the low spectral resolution of our data
precludes such measurements. 
The synthetic velocity profiles generated 
from the base models
are plotted in Figures 7 and 8, overlaid on the velocity centroid data for 
each galaxy.

We neglect any comparison in regions dominated by disk emission ($z\lesssim1$ 
kpc and $z\lesssim1.2$ kpc for NGC 891 and NGC 5775, respectively)
as the presence of dust in the disk prevents the sampling of
emission along the entire line of sight through each galaxy.  
As a result of this extinction, the
detected spectral lines are significantly biased towards outer disk emission 
and contain little emission from gas at the terminal velocity.  Due to 
sky-line contamination of the H$\alpha$ line in the spectra of 
NGC 5775, we use velocity centroids from [\ion{N}{2}] and [\ion{S}{2}] lines
in Figure 8.
We thus assume that the [\ion{N}{2}] and [\ion{S}{2}] lines trace the same 
radial distribution as H$\alpha$.  In the face-on galaxy M51 there is evidence
for a rise in [\ion{S}{2}]/H$\alpha$ with galactocentric radius (Greenawalt et
al. 1998) thus indicating the possibility that the [\ion{N}{2}] and 
[\ion{S}{2}] lines are biased towards outer disk emission in the spectra of 
edge-on galaxies.  However, in the case of NGC 891 at least, any 
radial dependence of line ratios has a negligible 
effect on observed velocities as the velocity centroids of 
[\ion{N}{2}]$\lambda6583$ and H$\alpha$ are fairly similar for the full run 
versus $z$.

It should also be noted that in these comparisons, it is impossible to know
the exact distribution of DIG along the line of sight.  DIG preferentially
concentrated in the outer disk can bias velocity centroids towards systemic.
For example, an artificially shallow gradient due to a decreased projection
of the rotation velocity vector along the line of sight 
can occur in a given slit if the DIG 
is concentrated in a filament in the outer disk.  We suspect 
this could be a 
problem mainly in NGC 5775 where DIG 
emission in the slit is primarily of a filamentary morphology and thus 
the distribution along the line of sight is not well known.  In the case
of NGC 891, the smoother appearance of the DIG layer tangential to the line of 
sight suggests the DIG is distributed fairly uniformly parallel to the line of
sight.   

\subsection{NGC 891}

The kinematic data for NGC 891 shown in Figure 7
exhibits a steady decrease in rotation speed with 
$z$ above 1 kpc.    The data indicates a drop in
$V_{sys}-V_{hel}$ of $\sim30$ km s$^{-1}$ from $z=1$ to 4.5 kpc on either side 
of the disk, while the ballistic model predicts a drop over the same range
in $z$ of $\sim85$ km s$^{-1}$, nearly equivalent to that of the fluid
disk model of Benjamin (2000) without the effects of pressure gradients
or magnetic fields.  The model with
$V_{k}=140$ km s$^{-1}$ in Figure 1 can explain the kinematics up to
$z=2$ kpc but fails beyond that height, and in any case does not match
the observed scale height of emission.  
 Clearly a ballistic model cannot explain
the vertical velocity gradient observed in NGC 891.  An inwardly directed
radial pressure gradient would only make the predicted velocity drop-off 
steeper, and is thus ruled out for explaining the rotation of the DIG halo.  
Thus a source of coupling between disk and halo rotation may be necessary to 
``speed-up'' the rotation of gas above the midplane.  One possibility is 
through viscous forces created during cloud collisions which our model 
ignores.  Magnetic effects should also play a role, given the dominant 
dynamical role of magnetic fields at high-$z$ (Boulares \& Cox 1990).
This could take the form of an outwardly directed magnetic 
pressure gradient, inwardly
directed magnetic tension, or a magnetic coupling between the disk and
halo. Models incorporating viscosity, magnetic tension, and pressure
gradients  would be extremely useful to 
determine if these effects significantly influence rotation. 

\subsection{NGC 5775}

NGC 5775 is at an inclination of $i=86\arcdeg$ and thus data for 
$z\lesssim1.2$ kpc represents emission from highly inclined disk 
structure (i.e. $z$ is the apparent height above the major axis).   
The data shown in Figure 8 for Slit 1 exhibits an asymmetry in $V_{hel}$ 
between the
NE and SW sides of the galaxy.  The sense of the asymmetry is consistent
with an outflow in the $z$-direction or, alternatively, a radial inflow similar
to that of the anomolous \ion{H}{1}\ component of NGC 2403 observed by
Fraternali et al. (2001), though such an inflow is not predicted in a 
fountain-like flow such as our ballistic model.
The asymmetry in the ballistic model is 
in the opposite sense as the non-circular line-of-sight velocities 
are dominated by {\it outward} radial migration.  
However,  emission on the NE side of Slit 1 
originates in a large filament which may trace
a chimney where peculiar kinematics may not be representative of halo rotation
as a whole. 

The gas kinematics for $z>1.2$ kpc in NGC 5775 seem to be better represented 
by the ballistic model than for NGC 891.  Though the exact velocities are 
slightly offset, the drop-off in $V_{sys}-V_{hel}$ from $z=1.2$ to 4.2 kpc for
the NE side of Slit 1 are similar for both the data and the model:
$\sim55$ km s$^{-1}$ and $\sim50$ km s$^{-1}$, respectively.  Again,
this region does trace the very bright DIG filament where unusual gas 
kinematics are likely.  The agreement
between data and model is not as good for the SW side over the same range in 
$z$, where the observed drop-off is $\sim20-30$ km s$^{-1}$.  
For Slit 2, the model can adequately reproduce some
of the kinematic data, particularly from $z=2$ to 4 kpc on either side of the 
disk, though the predicted velocity gradient of 65-70 km s$^{-1}$ from 
$z=1.2$ to 4.2 kpc is considerably steeper than indicated by the data for the
NE side.  Given the scatter of the data on the SW side, however, the 
observed gradient is generally consistent with the model prediction.   As 
for NGC 891, these cases of a  shallower observed velocity gradient than
predicted by the ballistic model suggest a mechanism of drag between 
disk and halo or an outwardly directed pressure gradient.  
The model with $V_{k}=110$ km s$^{-1}$ in Figure 2 does a 
somewhat better job in explaining the gas kinematics up to $z=4$ kpc, though
the vertical distribution of clouds in that model does not match the observed
emission scale height.

It should be noted that the ballistic model, including any of the variations
on the base model, completely fails for $z\gtrsim5$
kpc.  In each case, except for the SW side of Slit 1,  the data indicate 
that the gas velocities nearly reach systemic. {\it In no case do we ever find 
a ballistic model which produces zero velocity with respect to the underlying disk. Gas always shows some evidence of the underlying rotation.}
The difference between the data and model is not likely to be explained 
solely by a projection effect such as a 
concentration of the gas in the outer disk.  In fact, an analysis of the 
ballistic model of NGC 5775 shows that for $z>5$ kpc, all of the clouds 
reside at galactocentric radius of $R>8$ kpc (clouds at higher galactocentric
radius can reach
higher in $z$ due to a weaker potential at larger $R$).  Thus these clouds 
are well into
the outer disk for lines of sight through the model at positions determined by
the slits for the spectra.  In order to explain the velocity 
centroids near systemic, a source of gas support must be included in the 
model such as an inwardly directed thermal or magnetic pressure gradient.  
Also, the companion galaxy, NGC 5774, may play a role as a tidal 
interaction could affect the high-$z$ gas kinematics.  

\section{Conclusions}

We have presented a ballistic cloud model of disk-halo cycling, 
and from this model,
generated synthetic mean velocity profiles versus $z$.  We have compared
these profiles to velocity centroid data for the edge-on galaxies NGC 891 and
NGC 5775.  Though simplistic, these models have shown that hydrodynamic and
possibly magnetohydrodynamic effects are important for halo rotation.  A 
better understanding of halo kinematics may also be important for interpreting
QSO absorption lines from higher redshift disk-halo systems (e.g. Steidel et 
al. 2002). Through this work we have made the following conclusions:

1. The vertical velocity gradient observed in NGC 891 is
not as steep as predicted by the ballistic cloud model.  This suggests the
presence of drag between disk and halo such as through magnetic tension 
or viscous interactions between clouds. Alternatively, an outwardly directed
pressure gradient could explain the gas kinematics.

2. The ballistic model is more successful in explaining DIG
kinematics in NGC 5775.  The filamentary morphology of some of the DIG 
emission in this galaxy, suggesting the presence of outflows, may explain why  
the ballistic model provides a reasonable representation.  However, further
simulations of chimneys in a rotating frame would be necessary to confirm 
this speculation.   The ballistic model 
completely fails at high-$z$ where velocities nearly reach systemic.
An inwardly directed pressure gradient may provide 
the extra support needed to explain the apparent slow rotation at high-$z$. 
A possible tidal effect due to its companion galaxy, NGC 5774, should be 
considered when interpreting the kinematics observed in NGC 5775. 

3. The ballistic model predicts that clouds migrate radially outward as they
cycle through the halo.  The mass fluxes estimated from the models of NGC 891
and NGC 5775 imply that significant amounts of gas can be involved in these 
migrations. 
  Such migrations could
cause a redistribution of gas that could affect metallicity gradients as well
as star formation properties. Such effects have been previously investigated by Charlton \& Salpeter (1989), for example, but extensive observations of the kinematic behavior of edge-on galaxies should yield important constraints on such redistribution. 

4. We are limited in this work by the fact that we have data for only a few
slit positions. Hence, we are beginning to carry out Fabry-Perot 
observations of DIG in edge-on galaxies which will allow us to obtain
kinematic information with full two-dimensional coverage.  Such work will
allow issues of halo rotation to be addressed more completely.

\acknowledgments

This work was partially supported by NSF grant AST-9986113 to R.J.R. and NASA Theory grant NAG 5-8417 to R.A.B.

\begin{figure}
\figurenum{1}
\plotone{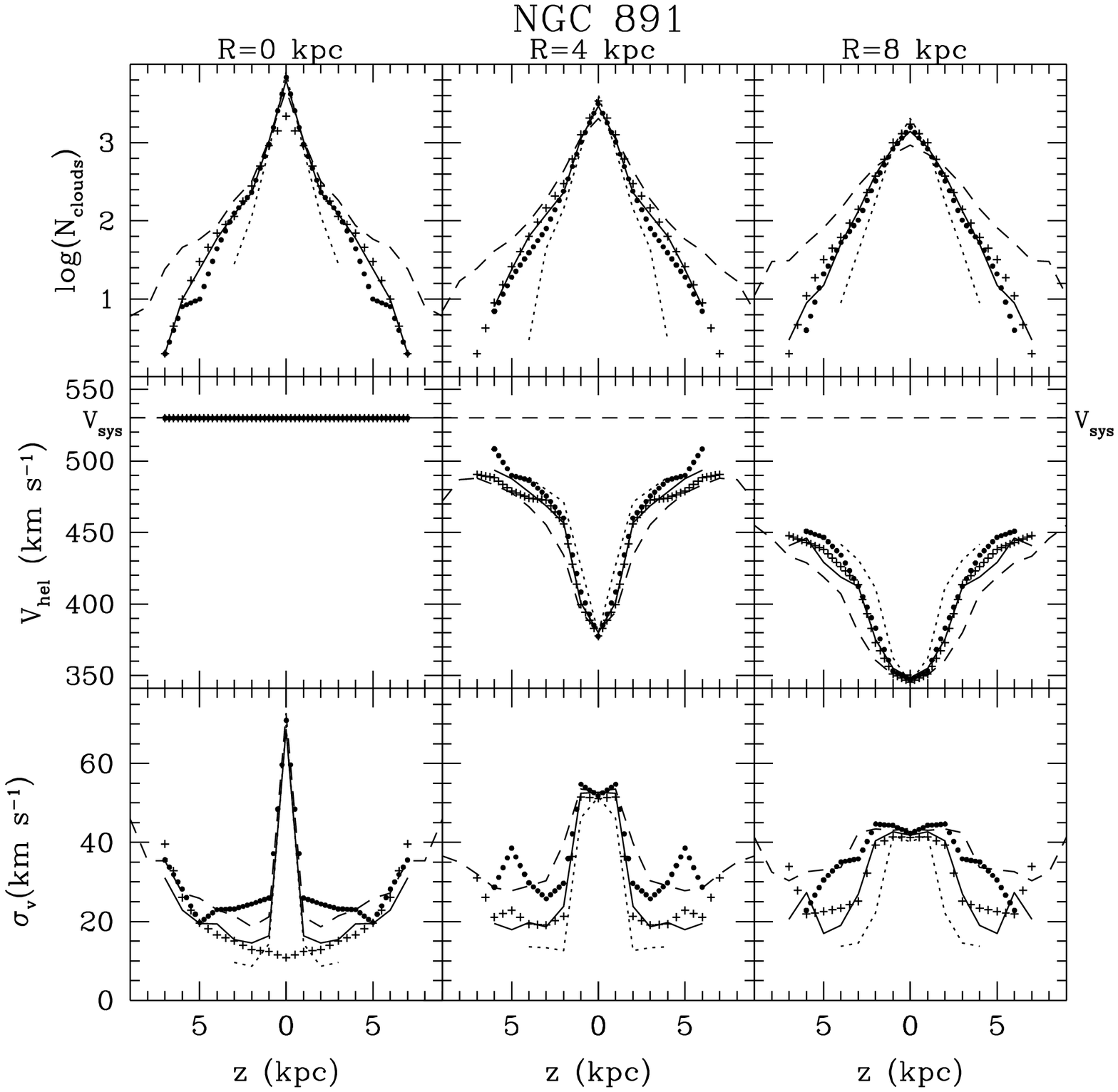}
\caption{The logarithm of number of clouds, $V_{hel}$, and $\sigma_{v}$ as
a function of height off the plane for a variety of models of NGC 891 at 
slit positions of $R=0$, 4, and 8 kpc on the approaching side of the
galaxy.  We consider a base model and various models with departures of one
parameter from this model (see Table 1).
The various models are:  base model (\em{solid line}\em), 
$V_{k}=70$ km s$^{-1} $(\em{small dashes}\em),
$V_{k}=140$ km s$^{-1}$ (\em{large dashes}\em), 
$\gamma_{0}=20\arcdeg$ (\em{dots}\em), 
and $R_{hole}=3.5$ kpc (\em{plus signs}\em).
The systemic velocity of $V_{sys}=530$ km s$^{-1}$ is indicated by the dashed 
line in the plots of $V_{hel}$.}
\end{figure}

\begin{figure}
\figurenum{2}
\plotone{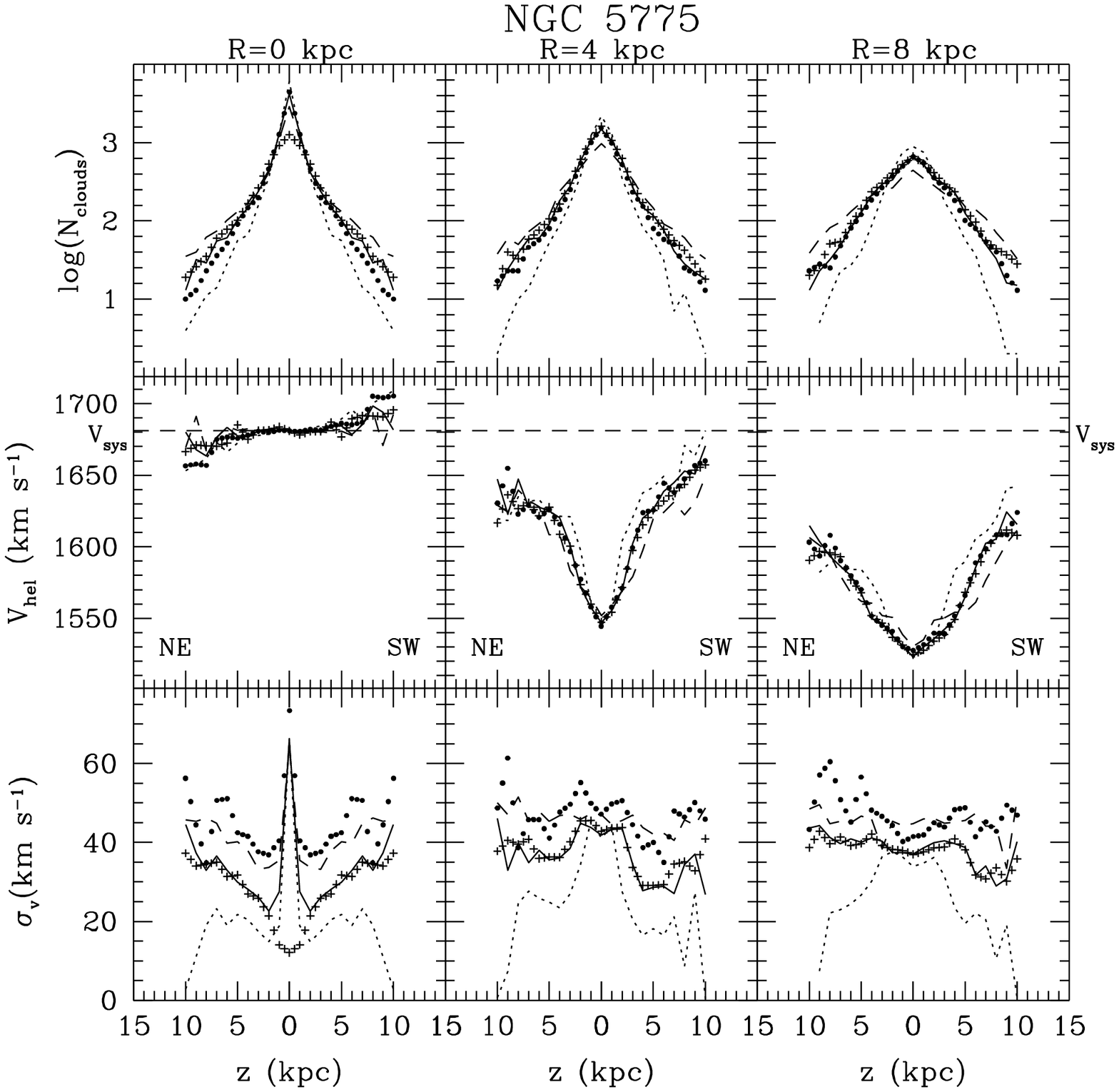}
\caption{The logarithm of number of clouds, $V_{hel}$, and $\sigma_{v}$ as
a function of height off the plane for a variety of models of NGC 5775 at 
slit positions of $R=0$, 4, and 8 kpc on the approaching side of the 
galaxy.  We consider a base model and various models with departures of one
parameter from this model (see Table 2).
The various models are:  base model (\em{solid line}\em), 
$V_{k}=110$ km s$^{-1} $(\em{small dashes}\em),
$V_{k}=210$ km s$^{-1}$ (\em{large dashes}\em), 
$\gamma_{0}=20\arcdeg$ (\em{dots}\em), 
and $R_{hole}=3$ kpc (\em{plus signs}\em).
The systemic velocity of $V_{sys}=1681$ 
km s$^{-1}$ is indicated by the dashed 
line in the plots of $V_{hel}$.}
\end{figure}


\begin{figure}
\figurenum{3}
\plotone{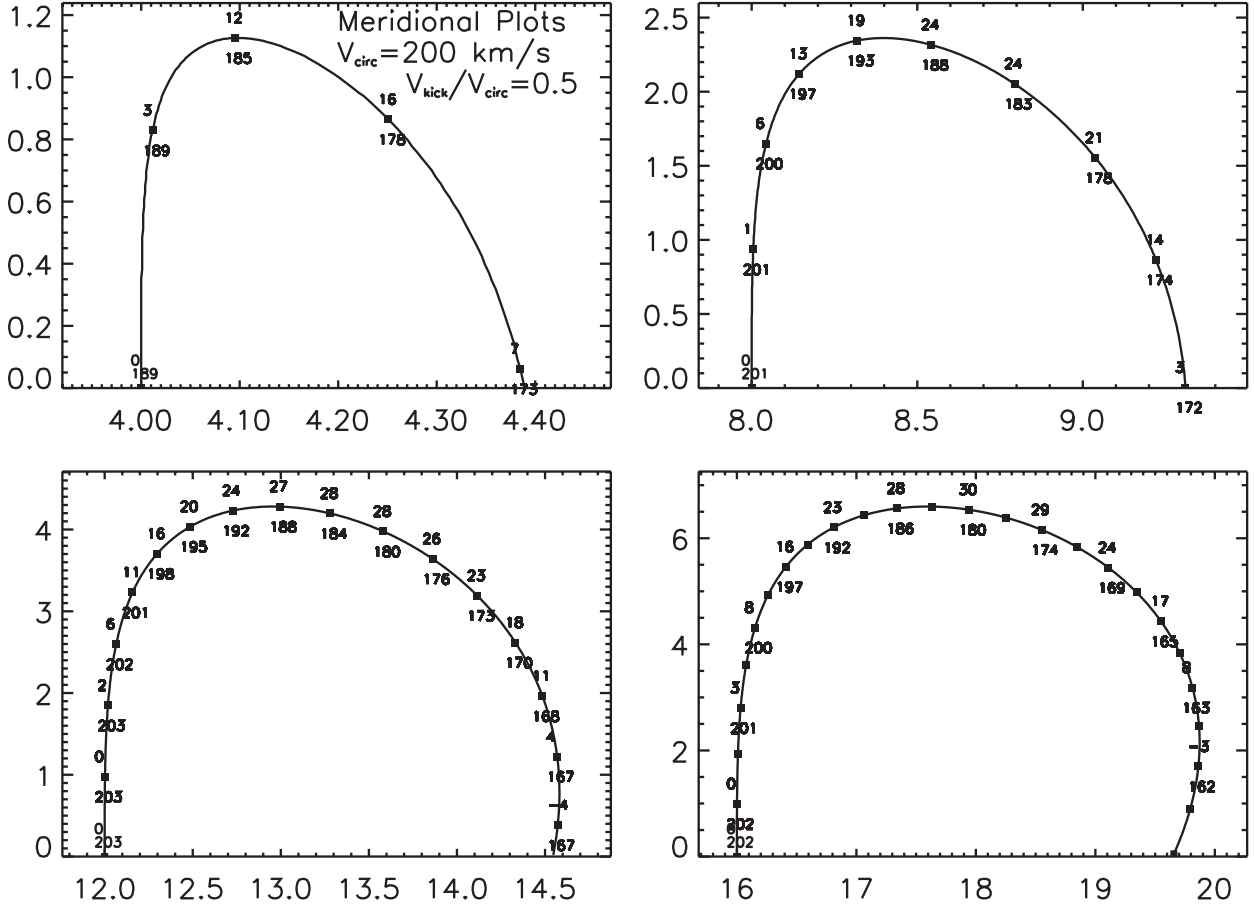}
\caption{Meridional plots showing sample orbits for a case with
$V_{c}=200
~{\rm km~s^{-1}}$ and $V_{k}=100~ {\rm km~s^{-1}}$ using the
gravitational
potential of Wolfire et al (1995). In each case, the horizontal axis is
radius in kiloparsecs, and the vertical axis is height in kiloparsecs.
These orbits start at
$R=4$ kpc [upper left], $R=8$ kpc [upper right], $R=12$ kpc [lower
left],
and $R=16$ kpc [lower right]. Solid points note the position of the
particle at 20 million year intervals. The number above each point
indicates the outward radial velocity; the number below gives the
azimuthal velocity.}
\end{figure}

\begin{figure}
\figurenum{4}
\plotone{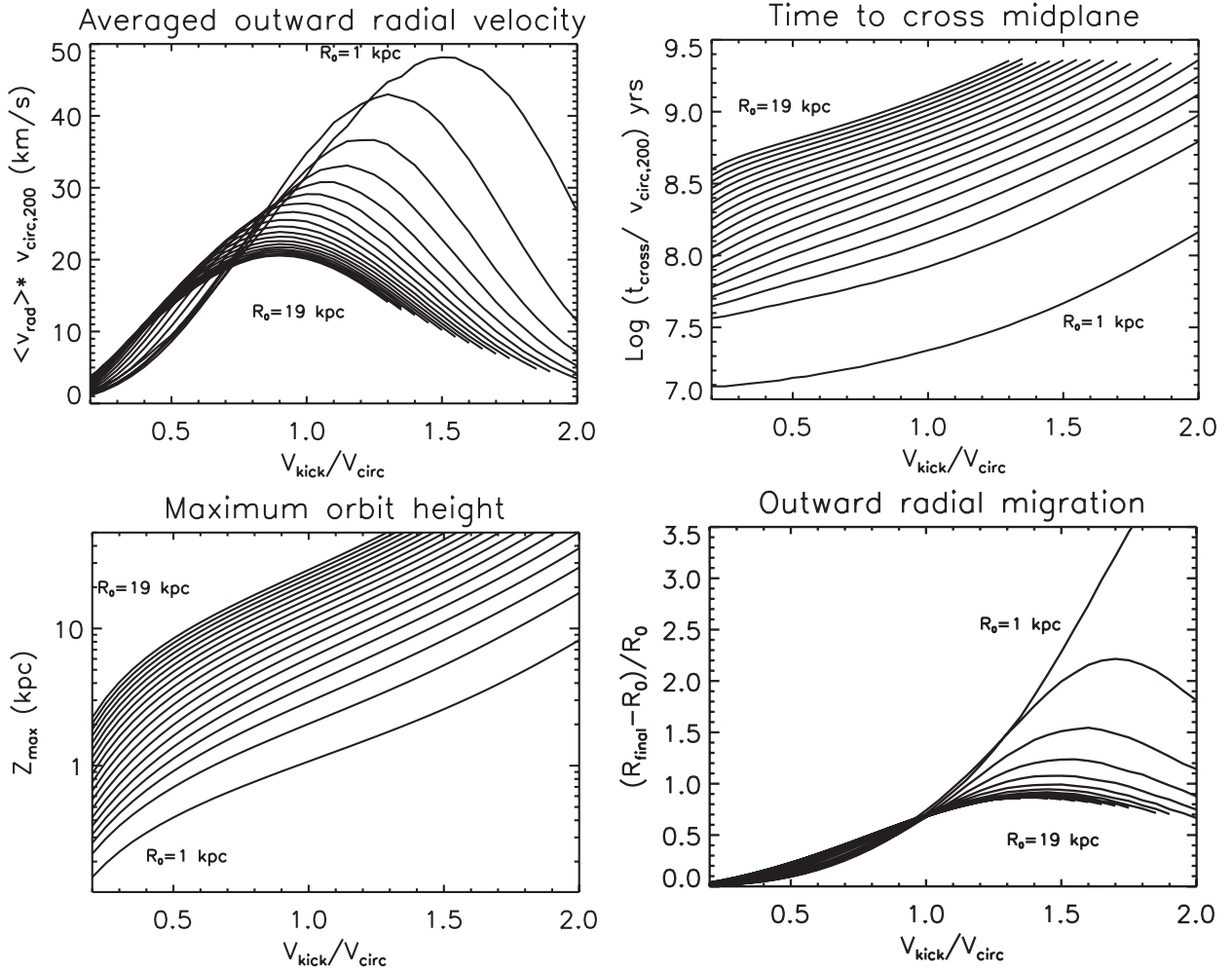}
\caption{Characteristics of orbits using the galactic potential model of
Wolfire et al. (1995). Each curve is for a different value of the
initial radius, and shows how characteristics vary as a function of the
ratio of the vertical kick velocity to the circular speed of the Galaxy,
$V_{k}/V_{c}$. [Top left] The average (outward) radial velocity of
orbits. Note that this velocity scales linearly with $V_{c}$ and is
normalized to
$V_{c}=200~ {\rm km~s^{-1}}$.  [Top right] The time it takes for a
particle to cross through the miplane. [Bottom left] Maximum height
obtained during each orbit. [Bottom right] The fractional change in
radius,
$(R_{final}-R_{0})/R_{0}$,
as a function of $V_{k}/V_{c}$. For $R_{o} > 6$ kpc, this
radial drift is relatively insensitive to radius.  }
\end{figure}

\begin{figure}
\figurenum{5}
\plotone{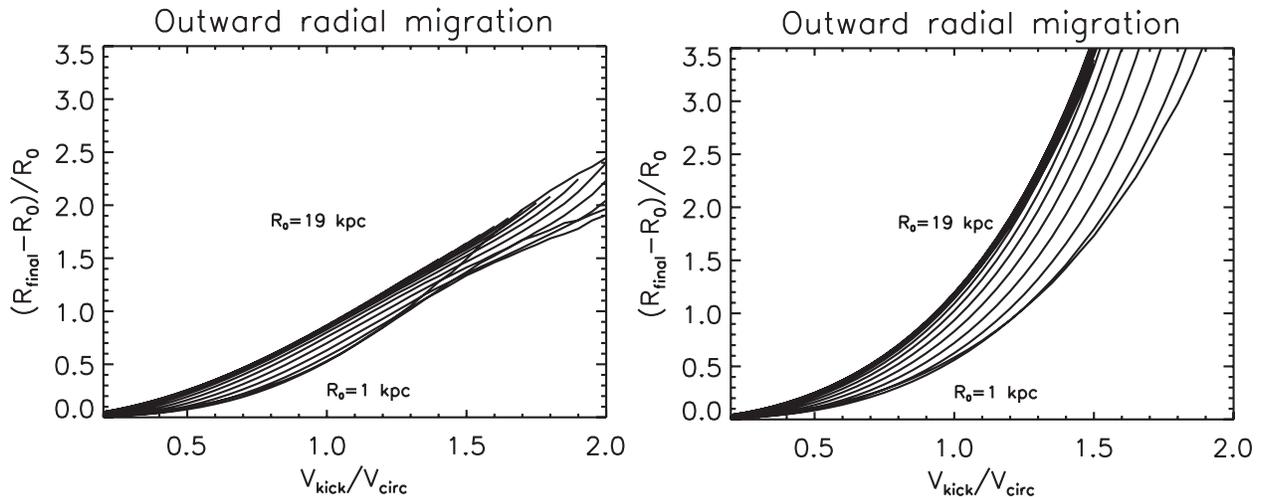}
\caption{The fractional radial change as a function of $V_{k}/V_{c}$ for
different initial radii using the galactic potential Model 2 [left
panel] and Model 2i [right panel] of Dehnen \& Binney (1998). Note
that for large values of $V_{k}/V_{c}$ the outward radial migration in
both cases is larger than that predicted by the Wolfire et al model
shown in the previous figure.  Model 2i uses a flattened halo
distribution with an axial ratio $q=0.3$, while Model 2 uses a rounder
halo with $q=0.8$. }
\end{figure}

\begin{figure}
\figurenum{6}
\plotone{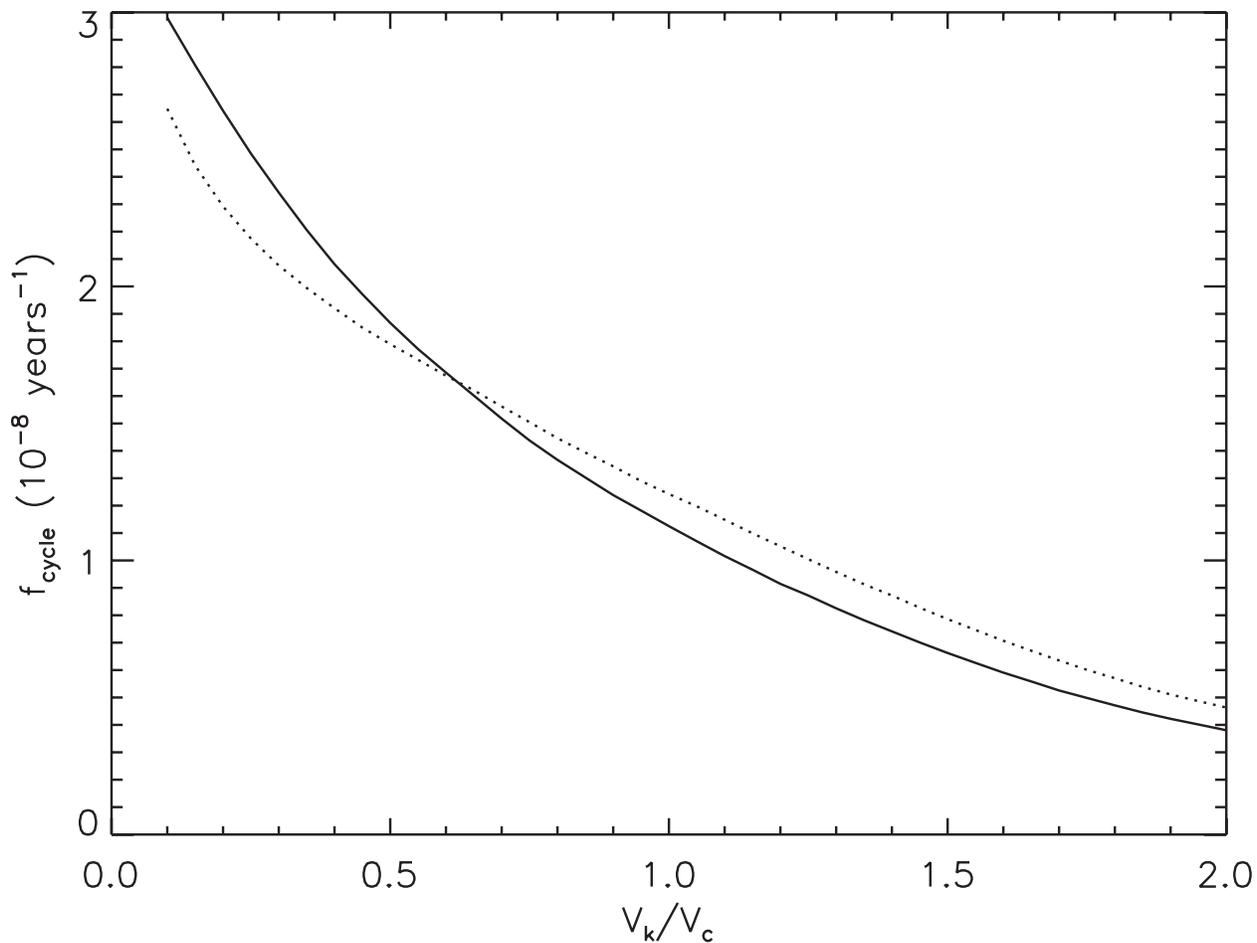}
\caption{The cycling frequency $f_{cycle}$ relating the halo mass flux
to the total halo mass (see text) as a function of $V_{k}/V_{c}$ using
the Wolfire et al (1995) potential [solid line] and the Dehnen \& Binney
(1998) potential [dotted line]. The initial orbit parameters
given in Section 2 are used.}
\end{figure}

\begin{figure}
\figurenum{7}
\plotone{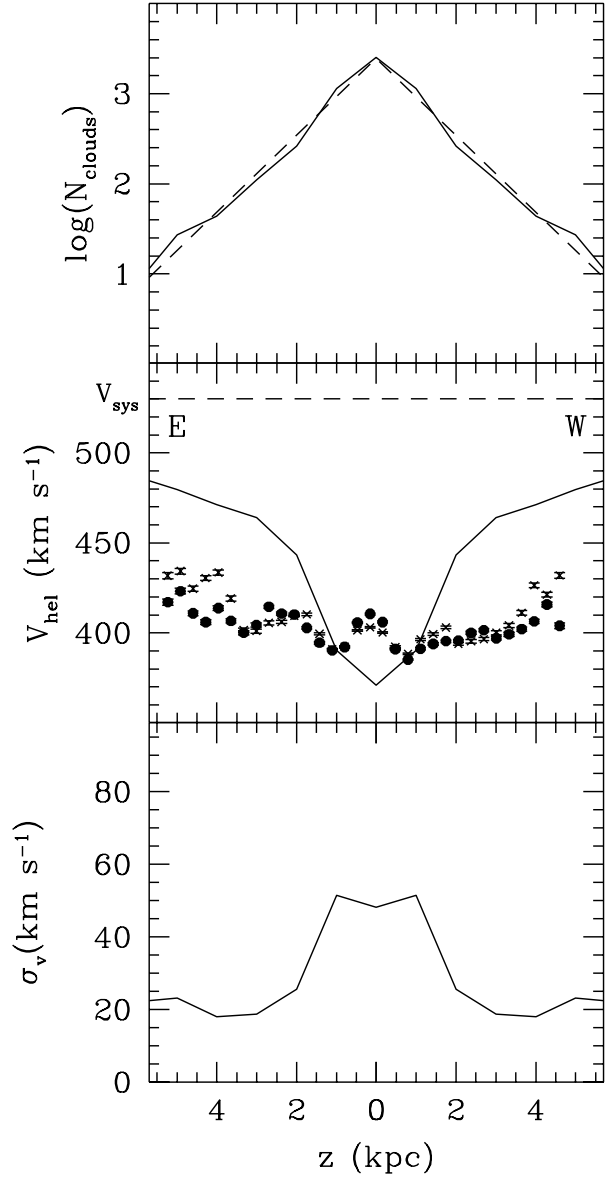}
\caption{The logarithm of number of clouds, $V_{hel}$, and $\sigma_{v}$ as
a function of height off the plane for the base model (\em{solid line}\em) 
of NGC 891 at the slit position of the 
observations ($R=4.6$ kpc from the galactic center on the approaching side).
In the top plot, the dashed line represents an exponential best-fit to the 
cloud distribution with scale height $h_{cl}=1$ kpc. 
Line centroid data from the H$\alpha$ 
($\times$) and [\ion{N}{2}]$\lambda$6583 ($\bullet$) are shown as well.  The
systemic velocity of $V_{sys}=530$ km s$^{-1}$ is indicated by the dashed 
line in the plot of $V_{hel}$.}
\end{figure}

\begin{figure}
\figurenum{8}
\plotone{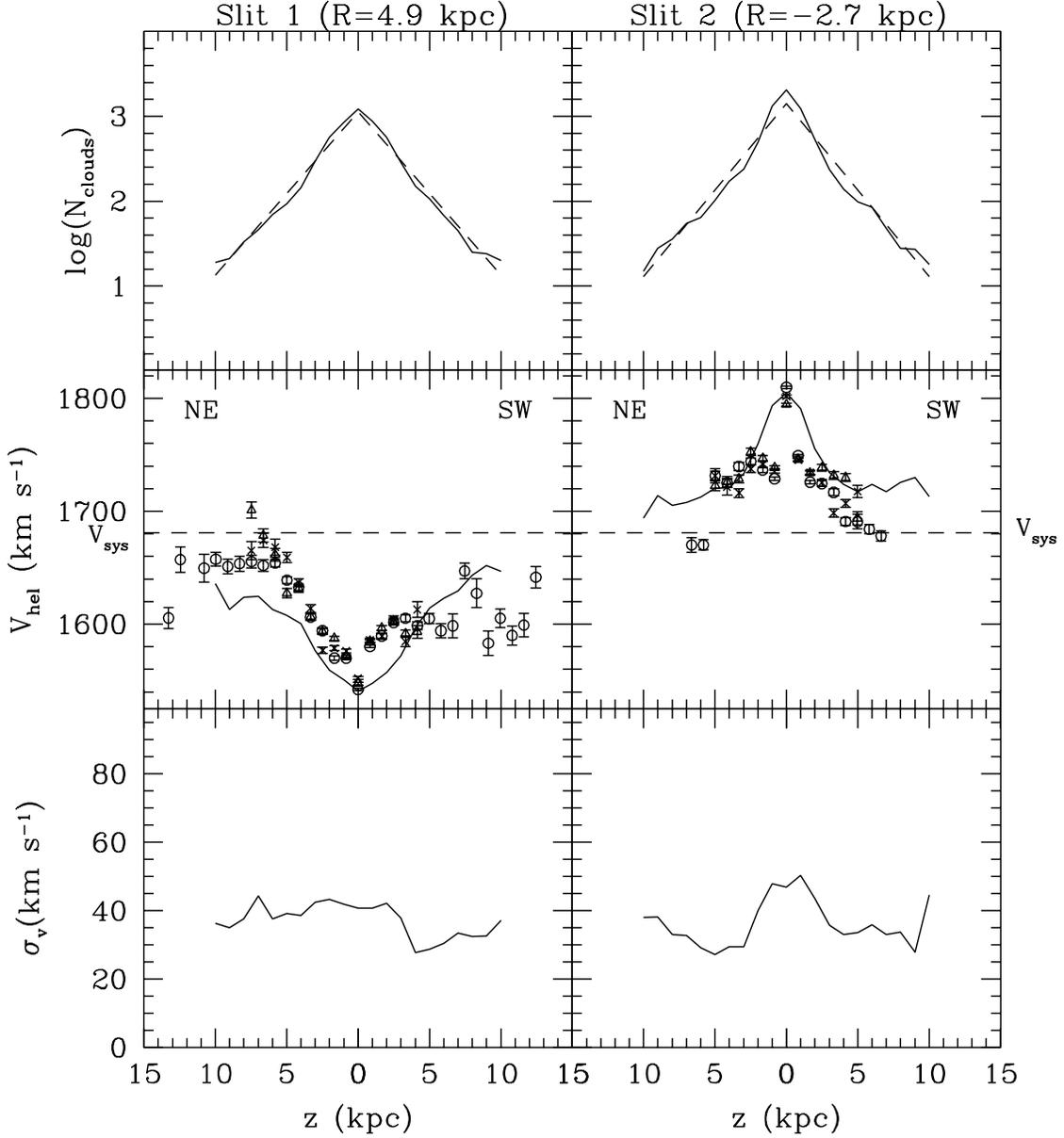}
\caption{The logarithm of number of clouds, $V_{hel}$, and $\sigma_{v}$ as
a function of height off the plane for the base model (\em{solid line}\em) 
of NGC 5775 at the slit positions of the 
observations ($R=4.9$ kpc from the galactic center on the approaching side and
$R=2.7$ kpc on the receding side for Slits 1 and 2, respectively).
In the top plots, the dashed lines represent the exponential best-fits to the 
cloud distribution with scale height $h_{cl}=2.2$ kpc and $h_{cl}=2.1$ kpc
for Slits 1 and 2, respectively.
Line centroid data from the [\ion{N}{2}]$\lambda$6583 ($\circ$), 
[\ion{S}{2}]$\lambda$6716 (\small{$\triangle$}), and [\ion{S}{2}]$\lambda$6731 
($\times$) lines are shown as well.  The systemic velocity of $V_{sys}=1681$ 
km s$^{-1}$ is indicated by the dashed 
line in the plots of $V_{hel}$.}
\end{figure}

\clearpage
\begin{deluxetable}{lccccccccc}
\tablecolumns{10}
\tablewidth{0pc}
\tablecaption{CHARACTERISTICS OF NGC 891 MODELS \label{t1}}
\tablehead{
\colhead{} & \multicolumn{3}{c}{Scale} & \multicolumn{3}{c}{$\vert\Delta V_{hel}\vert$ from} & \multicolumn{3}{c}{Range of} \\
\colhead{Model} & \multicolumn{3}{c}{Height (kpc)} & \multicolumn{3}{c}{$z$=0-3 kpc (km s$^{-1}$)} & \multicolumn{3}{c}{$\sigma_{v}$ (km s$^{-1}$)}  \\
\colhead{Description} & \colhead{$R$=0 kpc} & \colhead{$R$=4} & \colhead{$R$=8} &  \colhead{$R$=0} & \colhead{$R$=4} & \colhead{$R$=8} &
\colhead{$R$=0} & \colhead{$R$=4} & \colhead{$R$=8} }
\startdata
Base model\tablenotemark{a} & 1.0 & 1.0 & 1.1 & ... & 90  & 65 & 15-70 & 20-50 & 20-40 \\
$V_{k}=70$ km s$^{-1}$      & 0.5 & 0.6 & 0.7 & ... & 100 & 85 & 10-70 & 10-50 & 15-40 \\
$V_{k}=140$ km s$^{-1}$     & 1.5 & 1.5 & 1.8 & ... & 75  & 30 & 20-70 & 30-55 & 30-45 \\
$\gamma_{0}=20\arcdeg$      & 0.9 & 0.9 & 1.0 & ... & 95  & 65 & 20-70 & 25-55 & 30-45 \\
$R_{hole}=3.5$ kpc          & 1.1 & 1.0 & 1.1 & ... & 95  & 70 & 10-30 & 20-50 & 20-40 \\
\enddata
\tablenotetext{a}{The base model is characterized by the following parameters:
$R_{0}=7$ kpc, $V_{c}=230$ km s$^{-1}$, $V_{k}=100$ km s$^{-1}$, and 
$\gamma_{0}=0\arcdeg$.}
\end{deluxetable}

\clearpage
\begin{deluxetable}{lccccccccc}
\tablecolumns{10}
\tablewidth{0pc}
\tablecaption{CHARACTERISTICS OF NGC 5775 MODELS \label{t2}}
\tablehead{
\colhead{} & \multicolumn{3}{c}{Scale} & \multicolumn{3}{c}{$\vert\Delta V_{hel}\vert$ from} & \multicolumn{3}{c}{Range of} \\
\colhead{Model} & \multicolumn{3}{c}{Height (kpc)} & \multicolumn{3}{c}{$z$=0-3 kpc (km s$^{-1}$)} & \multicolumn{3}{c}{$\sigma_{v}$ (km s$^{-1}$)} \\
\colhead{Description} & \colhead{$R$=0 kpc} & \colhead{$R$=4} & \colhead{$R$=8} &  \colhead{$R$=0} & \colhead{$R$=4} & \colhead{$R$=8} &
\colhead{$R$=0} & \colhead{$R$=4} & \colhead{$R$=8}}
\startdata
Base model\tablenotemark{a} & 2.0 & 2.1 & 2.5 & ... & 55 & 20 & 25-65 & 25-45 & 30-40 \\
$V_{k}=110$ km s$^{-1}$     & 1.3 & 1.5 & 1.5 & ... & 75 & 35 & 15-65 & 15-45 & 20-35  \\
$V_{k}=210$ km s$^{-1}$     & 2.5 & 2.9 & 4.2 & ... & 30 & 20 & 35-65 & 40-50 & 40-50 \\
$\gamma_{0}=20\arcdeg$      & 1.7 & 2.1 & 2.6 & ... & 55 & 15 & 35-75 & 40-55 & 40-55 \\
$R_{hole}=3$ kpc            & 2.3 & 2.2 & 2.8 & ... & 50 & 20 & 15-35 & 30-45 & 30-40  \\
\enddata
\tablenotetext{a}{The base model is characterized by the following parameters:
$R_{0}=6$ kpc, $V_{c}=198$ km s$^{-1}$, $V_{k}=160$ km s$^{-1}$, and 
$\gamma_{0}=0\arcdeg$.}
\end{deluxetable}

\clearpage
\begin{deluxetable}{lccc}
\tablecolumns{4}
\tablewidth{0pc}
\tablecaption{MODEL HALO PROPERTIES \label{t3}}
\tablehead{
\colhead{Galaxy} & \colhead{$M_{h}$ ($\sqrt{\frac{f_{V}}{0.2}}$ M$_{\sun}$)} & \colhead{$\dot{M_{h}}$ ($\sqrt{\frac{f_{V}}{0.2}}$ M$_{\sun}$ yr$^{-1}$)} & \colhead{$\dot{E}$ ($\sqrt{\frac{f_{V}}{0.2}}$ erg s$^{-1}$)}} 
\startdata
NGC 891 & $1.1\times10^{9}$ & 22 & $2.4\times10^{40}$ \\
NGC 5775 & $9.2\times10^{8}$ & 13 & $3.7\times10^{40}$ \\
\enddata
\end{deluxetable}

\end{document}